\documentclass[preprint,tightenlines,aps,showpacs]{revtex4}
\usepackage{amsmath}
\usepackage{amssymb}
\usepackage{epsfig}

\providecommand{\openone}{\leavemode\hbox{\small\Kern-3.5pt\normalsize1}}

\begin{document}

\title{$\pi B_{8}B_{8}$ and $\sigma B_{8}B_{8}$ couplings from a chiral
quark potential model}
\author{M.T. Fern\'andez-Caram\'es}
\affiliation{Departamento de F\' \i sica Te\'orica e IFIC, Universidad de Valencia -
CSIC, E-46100 Burjassot, Valencia, Spain}
\author{P. Gonz\'alez}
\affiliation{Departamento de F\' \i sica Te\'orica e IFIC, Universidad de Valencia -
CSIC, E-46100 Burjassot, Valencia, Spain}
\author{A. Valcarce}
\affiliation{Departamento de F\'\i sica Fundamental, Universidad de Salamanca, E-37008
Salamanca, Spain}
\date{\today}

\begin{abstract}
From an SU(2)$\otimes $SU(2) chiral quark potential model incorporating
spontaneous chiral symmetry breaking the asymptotic $\pi $ and $\sigma $
exchange pieces of the $NN$ potential are generated. From them the $\pi NN$
and $\sigma NN$ coupling constants can be extracted. The generalization to
SU(3)$\otimes $SU(3) allows for a determination of $\pi B_{8}B_{8}$ and $%
\sigma B_{8}B_{8}$ coupling constants according to exact SU(3) hadron
symmetry. The implementation of the values of the couplings at $Q^{2}=0$
provided by QCD sum rules and/or phenomenology makes also feasible the
extraction of the meson-baryon-baryon form factors. In this manner a quite
complete knowledge of the couplings may be attained.
\end{abstract}

\pacs{12.39.Jh, 14.20.-c, 14.20.Gk}
\maketitle

\section{Introduction}

The meson-baryon-baryon $(mBB^{\prime })$ couplings play a central role in
hadron physics concerning the baryon-baryon $(BB^{\prime })$ interactions as
well as the formation and decay of baryon resonances. To study these
couplings effective hadron lagrangians involving the mesons and baryons
under consideration are postulated. All the complexity of the $mBB^{\prime }$
vertices is assumed to be taken into account through running couplings\
depending on $Q^{2}$, the transfer momentum in the vertex. This dependency
is usually parametrized in terms of a form factor and a coupling constant
defined as the value of the running coupling at a particular $Q^{2}$,
usually on-shell $Q^{2}=M_{m}^{2}$. Then one can calculate physical
processes and compare to data to extract these values. Thus the $\pi NN$
coupling constant is obtained from $\pi N$ and/or $NN$ scattering data~\cite%
{Eri92}. For couplings involving baryons and/or mesons for which scattering
or decay data are not so complete\ or unavailable one can also rely on
symmetry to derive predictions, see for instance~\cite{Sto96}.

From a more fundamental point of view hadrons are made up of quarks. Hence
hadron structures and decays as well as hadron-hadron interactions should
come out from quark dynamics as dictated by QCD. Due to the technical
difficulty to achieve this objective at present quark models of hadrons
incorporating QCD-motivated symmetries and dynamics have been successfully
applied to generate the baryon-baryon interactions, and consistently the
baryon spectrum, in the light $(u,d)$~\cite{Val05} as well as in the light +
strange $(u,d,s)$ quark sectors~\cite{Zha97,Shi00,Fuj07}. These models,
sometimes less precise than effective hadronic treatments, offer the
advantage of providing a consistent unified description of all baryon-baryon
processes from the same hamiltonian at the quark level. This confers them in
principle a great predictive power once the model parameters are tightly
constrained from some selected set of existing data.

We shall make use of this power to predict, within a non-relativistic chiral
quark model framework, $mB_{8}B_{8}$ ($B_{8}$: baryon of the flavor octet)
coupling constants in terms of meson-quark-quark ($mqq)$ couplings. More
precisely we shall generate, from a $mqq$ lagrangian incorporating the
effect of spontaneous chiral symmetry breaking (SCSB), the quark-quark meson
exchange potentials and from them, through a Born--Oppenheimer (BO)
approximation, the asymptotic baryon--baryon meson exchange interactions.
The use of justified harmonic oscillator baryon wave functions (in terms of
quarks) will allow us to perform analytic calculations. By comparing the
resulting interactions to the ones postulated at the effective hadronic
level we shall identify the meson-baryon-baryon coupling constants. This
procedure has been applied in the literature to the $\pi NN$ coupling~\cite%
{Liu93}. Here we shall be more precise in the extraction of coupling
constants and form factors and we shall extend its application to the $%
\sigma $ meson and the other $B_{8}$ baryons making feasible the comparison
of our results to the ones obtained with alternative methods based on quark
or hadron degrees of freedom.

The contents are organized as follows. In Sect.~\ref{secII} we shall center
on the light quark sector where spontaneously broken chiral SU(2)$\otimes $%
SU(2) symmetry serves as an underlying general guide to generate the
quark-quark meson-exchange potentials. We shall revisit the calculation of
the $\pi NN$ coupling in terms of the $\pi qq$ one and apply the same
procedure to the $\sigma NN$ case. We shall also comment on the possibility
of applying our method to $\Delta$ and nucleon resonances. Then in Sect.~\ref%
{secIII} we shall consider the extension, via chiral SU(3)$\otimes$SU(3), to
the SU(3) octet of baryons. Finally in Sect.~\ref{secIV} we shall summarize
our main results and conclusions.

\section{Light quark sector: SU(2)$\mathbf{\otimes}$SU(2).}

\label{secII} Our starting point is the chiral lagrangian 
\begin{equation}
\mathcal{L}_{ch}=-g_{ch}\,\bar{q}\,(\openone_{SU(2)}\sigma
\,+\,i\,\gamma _{5}\vec{\tau}\vec{\pi})q\,,  \label{eq2}
\end{equation}%
where $q$ has components $u$ and $d$, and $g_{ch}$ is the chiral $mqq$
coupling constant ($m:\pi ,\sigma $). The spontaneous chiral symmetry
breaking gives rise to a vertex form
factor $F(Q^{2})$ \cite{Dia86}. From the parametrization given in references~%
\cite{Fer86,Kus91} we propose the Lorentz invariant form (note that for the
purpose of the derivation of static potentials $Q^{2}=-\vec{Q}^{2}$), 
\begin{equation}
F(Q^{2})=\left( \frac{\Lambda ^{2}}{\Lambda ^{2}-Q^{2}}\right) ^{1/2}\,,
\end{equation}%
where $\Lambda $ is an effective cutoff parameter fitted from data. We
should keep in mind that this parametrization of $F(Q^{2})$ makes only sense
for $Q^{2}<\Lambda ^{2}.$

From the form proposed ($F(Q^{2}=0)=1)$ it is clear that $g_{ch}$
represents the value of the $mqq$ coupling, $g_{ch}F(Q^{2}),$ at $%
Q^{2}=M_{m}^{2}$ in the limit $M_{m}^{2}=0.$ In order to deal with coupling
constants defined on the physical meson masses ($M_{\pi }^{2}\neq 0\neq
M_{\sigma }^{2})$ we identify (it is implicitly assumed that $M_{\pi
}^{2},M_{\sigma }^{2}<\Lambda ^{2})$ 
\begin{equation}
g_{\pi qq}\equiv g_{ch}\left( \frac{\Lambda ^{2}}{\Lambda ^{2}-M_{\pi }^{2}}%
\right) ^{1/2}
\end{equation}%
\begin{equation}
g_{\sigma qq}\equiv g_{ch}\left( \frac{\Lambda ^{2}}{\Lambda ^{2}-M_{\sigma
}^{2}}\right) ^{1/2}\,,
\end{equation}%
as the values of the coupling $g_{ch}F(Q^{2})$ at $Q^{2}=M_{\pi
}^{2}$ and $Q^{2}=M_{\sigma }^{2}$, respectively.

Let us point out that the use of the experimental pion mass may have
required the introduction of an explicit symmetry breaking term proportional
to $\sigma $ in the lagrangian but this term has not any further effect in
the analysis we perform. Regarding the $\sigma $ mass it will be taken as a
parameter to be fitted from data around the value provided by the SCSB
relation~\cite{Sca82}, 
\begin{equation}
M_{\sigma }^{2}-M_{\pi }^{2}=4M_{q}^{2}\,
\end{equation}%
where $M_{q}$ denotes the constituent quark mass at $Q^{2}=0.$

In terms of $g_{\pi qq}$ we write the pion lagrangian as%
\begin{equation}
\mathcal{L}_{\pi qq}=-g_{\pi qq}\text{ }\bar{q}\,\,i\gamma _{5}\vec{\tau}%
\vec{\pi}\,\,q\,
\end{equation}%
with a vertex form factor $F_{\pi qq}(Q^{2})$ given by 
\begin{equation}
F_{\pi qq}(Q^{2})\equiv \left( \frac{\Lambda ^{2}-M_{\pi }^{2}}{\Lambda
^{2}-Q^{2}}\right) ^{1/2}\, .
\end{equation}%
Analogously, the sigma lagrangian reads 
\begin{equation}
\mathcal{L}_{\sigma qq}=-g_{\sigma qq}\bar{q}\,\,\openone_{SU(2)}\sigma \,\,q
\end{equation}%
with a vertex form factor $F_{\sigma qq}(Q^{2})$ given by 
\begin{equation}
F_{\sigma qq}(Q^{2})\equiv \left( \frac{\Lambda ^{2}-M_{\sigma }^{2}}{%
\Lambda ^{2}-Q^{2}}\right) ^{1/2} \, .
\end{equation}%
Note that both form factors are of the same type $F_{mqq}(Q^{2})\equiv \left(
\Lambda ^{2}-M_{m}^{2}/\Lambda ^{2}-Q^{2}\right) ^{1/2}$ so that $%
F_{mqq}(Q^{2}=M_{m}^{2})=1.$

\subsection{$mqq$ and induced $mBB$ potentials}

\label{secIIA}

From $\mathcal{L}_{mqq}$ and the form factors the static OPE and OSE central
potentials (to the respective lowest order in $Q^{2})$ can be obtained
through a non-relativistic reduction of the corresponding Feynman diagram
amplitudes. They are 
\begin{equation}
V_{\mathrm{OPE}}^{ij}({\vec{r}}_{ij})={\frac{1}{3}}{\frac{g_{\pi qq}^{2}}{{%
4\pi }}}\,{\ \frac{M_{\pi }^{2}}{{4M_{i}M_{j}}}}\,\,M_{\pi }\,\left[
\,Y(M_{\pi }\,r_{ij})-{\ \frac{\Lambda ^{3}}{M_{\pi }^{3}}}\,Y(\Lambda
\,r_{ij})\right] ({\vec{\sigma}}_{i}\cdot {\vec{\sigma}}_{j})({\vec{\tau}}%
_{i}\cdot {\vec{\tau}}_{j})\,,  \label{OPE}
\end{equation}%
\begin{equation}
V_{\mathrm{OSE}}^{ij}({\vec{r}}_{ij})=-{\frac{g_{\sigma qq}^{2}}{{4\pi }}}%
\,\,M_{\sigma }\,\left[ Y(M_{\sigma }\,r_{ij})-{\frac{\Lambda }{{M_{\sigma }}%
}}\,Y(\Lambda \,r_{ij})\right] \,.  \label{OSE}
\end{equation}%
Here $i$ and $j$ are numbers denoting quarks, $M_{i,j}=M_{q}$ , $\vec{\sigma}%
_{i,j}$ ($\vec{\tau}_{i,j})$ are the spin (isospin) Pauli operators, $r_{ij}$
is the interquark distance and the function $Y$ is defined as, 
\begin{equation}
Y(x)\,=\,{\frac{e^{-x}}{x}}\,.
\end{equation}%
Once the potentials at the quark level have been derived we shall use them
to obtain the baryon-baryon potentials. From $V_{\mathrm{OmE}}^{ij}$ the
asymptotic baryon-baryon meson exchange static potential is defined as 
\begin{equation}
(V_{q})_{\mathrm{OmE}}^{B_{a}B_{b}\rightarrow B_{c}B_{d}}(R\rightarrow
\infty )\equiv \lim_{R\rightarrow \infty }<\Psi _{B_{c}B_{d}}|\sum 
_{\substack{ i\in B_{a},B_{c}  \\ j\in B_{b},B_{d}}}V_{OmE}^{ij}|\Psi
_{B_{a}B_{b}}>\,,  \label{eq11}
\end{equation}%
where $\Psi _{B_{i}B_{j}}$ stands for the two-baryon wave function, $R$ for
the interbaryon distance and the integration is over the quark coordinates.

We shall concentrate on $B_{a}B_{b}\rightarrow B_{c}B_{d}$ interactions
involving baryons with the same mass. Then the asymptotic two-baryon wave
function will be expressed in the center of mass system as 
\begin{equation}
\Psi _{B_{a}B_{b}}=\Phi _{B_{a}}(1,2,3;+\vec{R}/2)\,\,\Phi _{B_{b}}(4,5,6;-%
\vec{R}/2)
\end{equation}%
where $(1,2,3)$ and $(4,5,6)$ denote the quarks forming the baryons, $\pm 
\vec{R}/2$ the baryons position and 
\begin{equation}
\Phi _{B_{i}}=(\Phi _{B_{i}})_{\mathrm{spatial}}(\Phi _{B_{i}})_{\mathrm{%
spin-flavor}}(\Phi _{B_{i}})_{\mathrm{color}}
\end{equation}%
is the one-baryon wave function expressed as the direct product of its
spatial, spin--flavor and color parts. For the sake of simplicity the baryon
spatial wave function will be chosen of harmonic oscillator type%
\begin{equation}
(\Phi _{B})_{\mathrm{spatial}}(1,2,3;+{\vec{R}}/2)=\prod_{i=1}^{3}\left( 
\frac{1}{\pi b^{2}}\right) ^{\frac{3}{4}}\exp \left[ -\left( \vec{r}_{i}-%
\vec{R}/2\right) ^{2}/2b^{2}\right]  \label{eqfin}
\end{equation}%
with an harmonic oscillator parameter, $b$, related to the size of the
baryon.

\subsection{Parameters}

In order to fix the parameters at the quark level: $M_q$, $M_{\pi}$, $%
M_{\sigma }$, $g_{ch}$, $\Lambda $, and $b$, we shall rely on the efficient
description of $NN$ data provided by the Chiral Quark Cluster Model (CQCM) 
\cite{Val05}. Such model contains, apart from OPE and OSE potentials derived
from $\mathcal{L}_{mqq}$, a confinement plus a residual one-gluon exchange
(OGE) interactions. We should realize though that the precise fitting of the
parameters in the CQCM relies on a RGM calculation so that the two-baryon
wave function as well as the baryon-baryon potentials are different from the
ones obtained via the Born-Oppenheimer (BO) approach we follow, Eq.~(\ref%
{eq11}). We shall take this into account in an effective manner by keeping
the same values for $M_{q}$, $M_{\pi}$, $M_{\sigma }$, $\Lambda$, and $b$,
and fitting a new value for $g_{ch}$ to reproduce, with our BO approach, the
experimental value of the pion-nucleon-nucleon coupling constant (see next
section). The values of the parameters used henceforth are listed in Table~%
\ref{t1}. 
\begin{table}[h]
\caption{Quark model parameters~\protect\cite{Val05}.}
\label{t1}%
\begin{tabular}{@{}lllll}
\hline
$M_{q}$ & $b$ & $M_{\sigma }$ & $M_{\pi }$ & $\Lambda $ \\ 
(MeV) & (fm) & (fm$^{-1}$) & (fm$^{-1}$) & (fm$^{-1}$) \\ \hline
313 & 0.518 & 3.42 & 0.7 & 4.2 \\ \hline
\end{tabular}%
\end{table}

\subsection{$\mathbf{\protect\pi NN}$}

\label{secIIB} 
\begin{figure}[h]
\vspace*{-1cm}
\epsfig{file=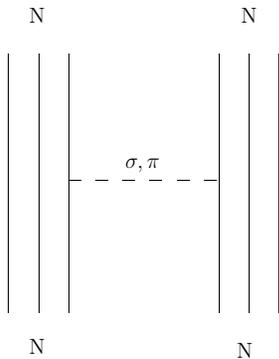,width=12cm}
\vspace*{-11cm}
\caption{Asymptotic $NN$ OPE or OSE interactions at quark level.}
\label{f1}
\end{figure}

From Eqs.~(\ref{OPE}), (\ref{eq11}) and (\ref{eqfin}) the
asymptotic OPE central potential for $NN\rightarrow NN$, corresponding to
the diagram of Fig.~\ref{f1} (there are nine equivalent ones), can be
obtained. It reads (the calculation has been explicitly done in Ref.~\cite%
{Liu93}) 
\begin{equation}
(V_{q})_{\mathrm{OPE}}^{NN\rightarrow NN}(R\rightarrow \infty )=\frac{g_{\pi
qq}^{2}}{4\pi }\left[ 9<(\vec{\sigma}_{3}\cdot \vec{\sigma}_{6})_{NN}.(\vec{%
\tau}_{3}\cdot \vec{\tau}_{6})_{NN}>e^{\frac{M_{\pi }^{2}b^{2}}{2}}\right] 
\frac{M_{\pi }^{2}}{4M_{q}^{2}}\frac{1}{3}\frac{e^{-M_{\pi }R}}{R}\,.
\label{eqa}
\end{equation}%
To extract the $\pi NN$ coupling we have to compare this potential with the
one derived from a postulated hadronic lagrangian. Assuming for instance a
pseudoscalar coupling we can write a lagrangian 
\begin{equation}
\mathcal{L}_{\pi NN}=-(g_{\pi NN})_{Q^{2}=M_{\pi }^{2}}\,\bar{N}%
\,\,i\,\gamma _{5}\vec{\tau}\,\vec{\pi}\,\,N\,,  \label{eqw}
\end{equation}%
with a vertex form factor $G_{\pi NN}(Q^{2})$ so that $G_{\pi
NN}(Q^{2}=M_{\pi }^{2})=1$. Notice that we have indicated explicitly the
on-shell character of the coupling $g_{\pi NN}$ through the subindex. In
order to derive from this lagrangian a Yukawa-like pion exchange potential,
monopole or dipole type form factors are usually assumed. Concerning the
asymptotic potential both give the same result. We shall use in parallel
with the form factor at the quark level a form 
\begin{equation}
G_{\pi NN}(Q^{2})\equiv \left( \frac{\Lambda _{\pi NN}^{2}-M_{\pi }^{2}}{%
\Lambda _{\pi NN}^{2}-Q^{2}}\right) ^{\frac{1}{2}}\,,  \label{pin2}
\end{equation}%
valid for $Q^{2}<\Lambda _{\pi NN}^{2}$ and $M_{\pi }^{2}<\Lambda _{\pi
NN}^{2}$ ($\Lambda _{\pi NN}$ is a cutoff parameter to be fitted).

Note that if we had preferred to refer the lagrangian to the value of the
coupling at $Q^{2}=0,$ i.e., to $(g_{\pi NN})_{Q^{2}=0}$ , then we would
have a different form factor such that 
\begin{equation}
(g_{\pi NN})_{Q^{2}=M_{\pi }^{2}}\,\,G_{\pi NN}(Q^{2})=(g_{\pi
NN})_{Q^{2}=0}\,\,\left( \frac{\Lambda _{\pi NN}^{2}}{\Lambda _{\pi
NN}^{2}-Q^{2}}\right) ^{\frac{1}{2}}\,,  \label{norm0}
\end{equation}%
where the second term on the right hand side represents the form factor
normalized at $Q^{2}=0.$

\begin{figure}[h]
\vspace*{-1cm}
\epsfig{file=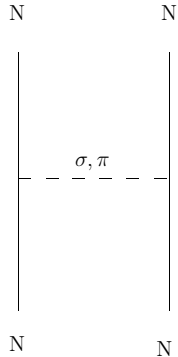,width=12cm}
\vspace*{-11cm}
\caption{Asymptotic $NN$ OPE or OSE interactions at baryon level.}
\label{f2}
\end{figure}

From $\mathcal{L}_{\pi NN}$ and $G_{\pi NN}(Q^{2})$ the non-relativistic
reduction of the one-pion exchange diagram$,$ Fig.~\ref{f2}, to the lowest
order in $Q^{2}$, provides us with the pion exchange static potential at the
baryonic level. The asymptotic behavior of its central part is given by%
\begin{equation}
(\mathcal{V}_{B})_{\mathrm{OPE}}^{NN\rightarrow NN}(R\rightarrow \infty )=%
\frac{(g_{\pi NN}^{2})_{Q^{2}=M_{\pi }^{2}}}{4\pi }\left[ \left\langle (\vec{%
\sigma}_{N}\cdot \vec{\sigma}_{N})(\vec{\tau}_{N}\cdot \vec{\tau}%
_{N})\right\rangle \right] \frac{M_{\pi }^{2}}{4M_{N}^{2}}\,\,\frac{1}{3}%
\frac{e^{-M_{\pi }R}}{R}\,.  \label{eqb}
\end{equation}%
Note that $(\mathcal{V}_{B})_{\mathrm{OPE}}^{NN\rightarrow NN}(R\rightarrow
\infty )$ only depends on the coupling constant and not on the form factor.
Then no information on the form factors at the baryon level can be extracted
from it. Regarding the coupling constant we can make use of the relation $%
(25/9)<(\vec{\sigma}_{N}\cdot \vec{\sigma}_{N})(\vec{\tau}_{N}\cdot \vec{\tau%
}_{N})>=9<(\vec{\sigma}_{3}\cdot \vec{\sigma}_{6})_{NN}.(\vec{\tau}_{3}\cdot 
\vec{\tau}_{6})_{NN}>$ \cite{Liu93} to compare Eqs.~(\ref{eqa}) and (\ref%
{eqb}). From this comparison we extract 
\begin{equation}
(g_{\pi NN}^{2})_{Q^{2}=M_{\pi }^{2}}=g_{\pi qq}^{2}\,\,\frac{M_{N}^{2}}{%
M_{q}^{2}}\,\,\frac{25}{9}\,\,e^{\frac{M_{\pi }^{2}b^{2}}{2}}\,.  \label{eqc}
\end{equation}%
Having chosen $M_{q}=313$ MeV so that $M_{N}=3M_{q}$ we can re-express 
\begin{equation}
(g_{\pi NN}^{2})_{Q^{2}=M_{\pi }^{2}}=25\,\,g_{\pi qq}^{2}\,\,e^{\frac{%
M_{\pi }^{2}b^{2}}{2}}\,.  \label{pinpiq}
\end{equation}

\subsubsection{Pseudovector couplings}

Alternatively to $g_{\pi qq}$ and $g_{\pi NN}$ we could have used
pseudovector couplings $f_{\pi qq}$ and $f_{\pi NN}$ so that 
\begin{equation}
\mathcal{L}_{\pi qq}^{PV}=-\frac{f_{\pi qq}}{M_{\pi }}\bar{q}\,\,i\gamma
_{5}\gamma _{\mu }\vec{\tau}\partial ^{\mu }\vec{\pi}\,\,q  \label{piqPV}
\end{equation}%
with the vertex form factor $F_{\pi qq}(Q^{2})$ and

\begin{equation}
\mathcal{L}_{\pi NN}^{PV}=-\frac{(f_{\pi NN})_{Q^{2}=M_{\pi }^{2}}}{M_{\pi }}%
\,\bar{N}\,\,i\,\gamma _{5}\gamma _{\mu }\vec{\tau}\partial ^{\mu }\vec{\pi}%
\,\,N\,.  \label{pinPV}
\end{equation}%
with the vertex form factor $G_{\pi NN}(Q^{2}).$

\noindent 
It turns out that $\mathcal{L}^{PV}$ give rise to exactly the same
potentials as $\mathcal{L}$ , to the lowest $Q^{2}$ order, under the
identifications 
\begin{equation}
\frac{f_{\pi qq}^{2}}{M_{\pi }^{2}}=\frac{g_{\pi qq}^{2}}{4M_{q}^{2}}
\end{equation}%
\begin{equation}
\frac{(f_{\pi NN}^{2})_{Q^{2}=M_{\pi }^{2}}}{M_{\pi }^{2}}=\frac{(g_{\pi
NN}^{2})_{Q^{2}=M_{\pi }^{2}}}{4M_{N}^{2}}\,.
\end{equation}%
If we now substitute these relations in Eq.~(\ref{eqc}) we get 
\begin{equation}
(f_{\pi NN}^{2})_{Q^{2}=M_{\pi }^{2}}=f_{\pi qq}^{2}\,\,\frac{25}{9}\,\,e^{%
\frac{M_{\pi }^{2}b^{2}}{2}}\,.  \label{eqt}
\end{equation}%
The corresponding constant $f_{ch}$ is consistently defined as 
\begin{equation}
f_{ch}=\left( \frac{\Lambda ^{2}-M_{\pi }^{2}}{\Lambda ^{2}}\right)
^{1/2}f_{\pi qq}=\lim_{M_{\pi }\rightarrow 0}f_{\pi qq}\,.
\end{equation}

\subsubsection{Coupling constants and form factor values}

From a standard experimental value $(f_{\pi NN}^{2})_{Q^{2}=M_{\pi
}^{2}}/4\pi \simeq 0.079$ (see~\cite{Eri92} and references therein) or $%
(g_{\pi NN}^{2})_{Q^{2}=M_{\pi }^{2}}/4\pi \simeq 14.6$ we fit the
pion-quark-quark coupling constant 
\begin{equation}
f_{\pi qq}^{2}/4\pi =0.027
\end{equation}%
or 
\begin{equation}
g_{\pi qq}^{2}/4\pi =0.55\,,  \label{gpiq}
\end{equation}%
and 
\begin{equation}
|g_{ch}|\equiv |f_{ch}|\,\dfrac{2\,M_{q}}{M_{\pi }}=2.6\,.
\end{equation}
Let us emphasize that this value for $g_{ch}$ differs less than a
10\% from the one obtained via QCD sum rules (QCDSR) $(g_{ch})_{%
\mathrm{QCDSR}}\simeq 2.83$ \cite{Hwa96}.

Regarding $\Lambda _{\pi NN}$, the cutoff parameter, its range of values can
be estimated. In Ref.~\cite{Hen91} a fit to data was attained from a
gaussian form factor $e^{Q^{2}/\Lambda _{\mathrm{HM}}^{2}}$ with $\Lambda _{%
\mathrm{HM}}$ varying from $2.6$ to $4.2$ fm$^{-1}.$ This form factor is
normalized at $Q^{2}=0$. By requiring its low $Q^{2}$ behavior $%
(1+Q^{2}/\Lambda _{\mathrm{HM}}^{2}),$ to be the same than that of our form
factor normalized at $Q^{2}=0$ in Eq.~(\ref{norm0}), $(1+Q^{2}/2\Lambda
_{\pi NN}^{2})$, we get $\Lambda _{\mathrm{HM}}^{2}=2\Lambda _{\pi NN}^{2}$.
Hence the resulting range for $\Lambda _{\pi NN}$ is 
\begin{equation}
\Lambda _{\pi NN}=1.84-2.97\,\,\,\mathrm{fm}^{-1}\,.
\end{equation}%
From Eqs.~(\ref{pin2}) and (\ref{norm0}) this range can be translated in an
interval of values for the coupling at $Q^{2}=0$ : 
\begin{eqnarray}
\frac{(f_{\pi NN}^{2})_{Q^{2}=0}}{4\pi } &=&0.068-0.075\,,  \notag \\
\frac{(g_{\pi NN}^{2})_{Q^{2}=0}}{4\pi } &=&12.5-13.8\,.
\end{eqnarray}%
These values are in perfect agreement with the phenomenological analysis
done in Ref.~\cite{Tho89} (let us comment that for the form factor
used in this reference the range of values for the cutoff parameter is the
same as for $\Lambda _{\mathrm{HM}}$). The preferred value in Ref.~\cite%
{Tho89} is $(f_{\pi NN}^{2})_{Q^{2}=0}/4\pi \simeq 0.073$ which corresponds
to $(g_{\pi NN}^{2})_{Q^{2}=0}/4\pi \simeq 13.5$. It is worthwhile to point
out that this corresponds quite approximately to the $M_{\pi }\rightarrow 0$
limit of our expression $(f_{\pi NN}^{2})_{Q^{2}=M_{\pi }^{2}}=f_{\pi
qq}^{2}\,(25/9)\,e^{\frac{M_{\pi }^{2}b^{2}}{2}}$, i.e., 
\begin{equation}
\dfrac{(f_{\pi NN}^{2})_{Q^{2}=M_{\pi }^{2}\rightarrow 0}}{4\pi }=\dfrac{f_{%
ch}^{2}}{4\pi }\frac{25}{9}=0.072  \label{gchiral}
\end{equation}%
indicating the quite approximate Goldstone boson character of the pion.

\subsection{$\mathbf{\protect\sigma NN}$}

\label{secIIC} By proceeding in exactly the same way for the $\sigma $
exchange we obtain from Eqs.~(\ref{OSE}), (\ref{eq11}) and (\ref{eqfin}) at
the quark level 
\begin{equation}
(V_{q})_{\mathrm{OSE}}^{NN\rightarrow NN}(R\rightarrow \infty )=-\frac{%
g_{\sigma qq}^{2}}{4\pi }\,\,9\,\,e^{\frac{M_{\sigma }^{2}b^{2}}{2}}\,\,%
\frac{e^{-M_{\sigma }R}}{R}\,.
\end{equation}%
On the other hand from the hadronic lagrangian 
\begin{equation}
\mathcal{L}_{\sigma NN}=-(g_{\sigma NN})_{Q^{2}=M_{\sigma }^{2}}\bar{N}\,\,%
\openone_{SU(2)}\sigma \,\,N\,,  \label{eqna}
\end{equation}%
with a vertex form factor 
\begin{equation}
G_{\sigma NN}(Q^{2})=\left( \frac{\Lambda _{\sigma NN}^{2}-M_{\sigma }^{2}}{%
\Lambda _{\sigma NN}^{2}-Q^{2}}\right) ^{\frac{1}{2}}\,,
\end{equation}%
we get 
\begin{equation}
(\mathcal{V}_{B})_{\mathrm{OSE}}^{NN\rightarrow NN}(R\rightarrow \infty )=-%
\frac{(g_{\sigma NN}^{2})_{Q^{2}=M_{\sigma }^{2}}}{4\pi }\,\,\frac{%
e^{-M_{\sigma }R}}{R}\,.
\end{equation}%
From their comparison 
\begin{equation}
(g_{\sigma NN}^{2})_{Q^{2}=M_{\sigma }^{2}}=9\,\,g_{\sigma qq}^{2}\,\,e^{%
\frac{M_{\sigma }^{2}b^{2}}{2}}\,.  \label{gsigq}
\end{equation}%
Let us note that once the value of $g_{\pi qq}$ has been fitted our model
predicts the value of $g_{\sigma qq}$ through 
\begin{equation}
\frac{g_{\sigma qq}^{2}}{g_{\pi qq}^{2}}=\frac{\Lambda ^{2}-M_{\pi }^{2}}{%
\Lambda ^{2}-M_{\sigma }^{2}}\,.  \label{eqk2}
\end{equation}%
Equivalently 
\begin{equation}
\frac{(g_{\sigma NN}^{2})_{Q^{2}=M_{\sigma }^{2}}}{(g_{\pi
NN}^{2})_{Q^{2}=M_{\pi }^{2}}}=\frac{9}{25}\,\,\frac{\Lambda ^{2}-M_{\pi
}^{2}}{\Lambda ^{2}-M_{\sigma }^{2}}\,\,e^{\frac{(M_{\sigma }^{2}-M_{\pi
}^{2})b^{2}}{2}}\,.  \label{eqk}
\end{equation}

\subsubsection{Coupling constant and form factor values}

From Eq.~(\ref{eqk}) we predict 
\begin{equation}
\frac{(g_{\sigma NN}^{2})_{Q^{2}=M_{\sigma }^{2}}}{4\pi }=68.2\,
\end{equation}%
and from Eq.~(\ref{gsigq})%
\begin{equation}
\frac{g_{\sigma qq}^{2}}{4\pi }=1.59\,.
\end{equation}
As mentioned above our asymptotic comparison does not give any information
on the cutoff parameter $\Lambda _{\sigma NN}$. Nonetheless we can combine
our result for the coupling constant with the value of the coupling at $%
Q^{2}=0$ provided by QCDSR to get an insight into it. From Ref.~\cite{Erk06} 
\begin{equation}
\dfrac{(g_{\sigma NN})_{Q^{2}=0}}{g_{q}^{\sigma }}=3.9\pm 1.0.
\end{equation}%
If we tentatively identify $g_{q}^{\sigma }$ with our $g_{\sigma qq}$ ($%
=4.47)$ we get 
\begin{equation}
\dfrac{\left( g_{\sigma NN}^{2}\right) _{Q^{2}=0}}{4\pi }=17.4\pm 4.5\,,
\label{gsiqcd}
\end{equation}
and using the relation $\left( g_{\sigma NN}^{2}\right) _{Q^{2}=0}=\left(
g_{\sigma NN}^{2}\right) _{Q^{2}=M_{\sigma }^{2}}\,G_{\sigma
NN}^{2}(Q^{2}=0) $ we extract 
\begin{equation}
\Lambda _{\sigma NN}=\left( \frac{M_{\sigma }^{2}}{1-\frac{(g_{\sigma
NN}^{2})_{Q^{2}=0}}{(g_{\sigma NN}^{2})_{Q^{2}=M_{\sigma }^{2}}}}\right) ^{%
\frac{1}{2}}=3.97\pm 0.18\,\,\mathrm{fm}^{-1} \, .  \label{cutoff}
\end{equation}%
It is again interesting to consider the limit $M_{\sigma }\rightarrow 0$ of
Eq.~(\ref{gsigq}) 
\begin{equation}
\dfrac{(g_{\sigma NN}^{2})_{Q^{2}=M_{\sigma }^{2}\rightarrow 0}}{4\pi }=%
\dfrac{9g_{ch}^{2}}{4\pi }=4.8\,,  \label{gsinchi}
\end{equation}%
or 
\begin{equation}
\dfrac{(g_{\sigma NN}^{2})_{Q^{2}=M_{\sigma }^{2}\rightarrow 0}}{g_{\mathrm{%
ch}}^{2}}=9
\end{equation}%
and compare it to the interval of values of the coupling at $Q^{2}=0$ from
Eq.~(\ref{gsiqcd}). As can be seen the $M_{\sigma }\rightarrow 0$ value from
Eq.~(\ref{gsinchi}) is out of this interval. This might be interpreted as
reflecting the non-Goldstone boson nature of the $\sigma .$

\subsection{$\mathbf{\protect\pi N\Delta}$}

\label{secIID} Strictly speaking our procedure to extract the couplings only
makes sense when the lowest order expansion in $Q^{2}$ we follow is
simultaneously valid at the baryonic level ($E_{B}\simeq M_{B})$ and at the
quark level ($E_{q}\simeq M_{q})$. We do not expect this to be true for
light-quark baryons in general since the quarks can move relativistically
inside them. However the structure of the ground states, the $N$ (and $%
\Lambda ,\Sigma ,$ and $\Xi $ when considering $SU(3)),$ is well described
by non-relativistic constituent quark models through the effective
parameters in the quark-quark potential. Then we expect our procedure to
make sense for them. Regarding other baryon states like $\Delta $ and $%
N(1440)$ one should be more cautious as we illustrate next.

In order to extract the $\pi N\Delta $ coupling constant we consider the $%
NN\rightarrow N\Delta $ interaction. According to the harmonic oscillator
model we are using the spatial wave function of $\Delta $ has exactly the
same structure than the $N$ one. However the real $\Delta $ differs from the 
$N.$ To implement the bigger size for $\Delta $ predicted by
non-relativistic spectroscopic models we shall consider the possibility of a
slightly different value for the size parameter. Thus the $\Delta $ spatial
wave function we shall use is 
\begin{equation}
(\Phi _{\Delta })_{\mathrm{spatial}}(4,5,6;+\vec{R}/2)=\prod_{i=4}^{6}\left( 
\frac{1}{\pi b_{\Delta }^{2}}\right) ^{\frac{3}{4}}\exp \left[ -{\left( \vec{%
r}_{i}-\vec{R}/2\right) ^{2}/2b_{\Delta }^{2}}\right]  \label{eqx}
\end{equation}%
with a baryon size parameter, $b_{\Delta }$. One should also keep in mind
that the mass of the $\Delta $ is a $30\%$ bigger than the mass of the $N$.
In our harmonic oscillator quark model this means that the quarks in the $%
\Delta $ have more potential and kinetic energy (virial theorem) than the
quarks in the $N$. According to our comments above this could give rise to
corrections in the expression of the static potential at the quark level.
Moreover due to the $\Delta -N$ mass difference the positions of the baryons
in the initial and final states should not be the same. Therefore we should
not expect an accurate prediction for the coupling constant in this case. 
\begin{figure}[h]
\vspace*{-1cm}
\epsfig{file=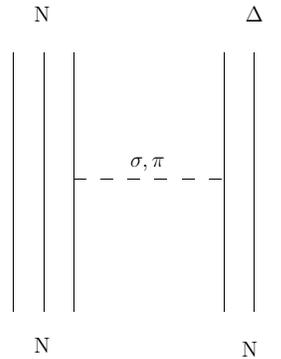,width=12cm}
\vspace*{-11cm}
\caption{Asymptotic $NN\rightarrow N\Delta $ OPE or OSE interactions at
quark level.}
\label{f3}
\end{figure}
To derive such prediction we first calculate at the quark level, from Eqs.~(%
\ref{eq11}), (\ref{eqfin}) and (\ref{eqx}) (we use for easiness the
pseudovector form of the coupling) the asymptotic $NN\rightarrow N\Delta $
pion exchange static central potential, Fig.~\ref{f3}. It is 
\begin{equation}
(V_{q})_{\mathrm{OPE}}^{NN\rightarrow N\Delta }(R\rightarrow \infty )=\frac{%
f_{\pi qq}^{2}}{4\pi }\left[ \frac{160}{9\sqrt{2}}\,\,2\sqrt{2}\,\,\frac{%
b_{\mu }^{6}b_{+}^{3}b_{e}^{3}}{(b^{15}b_{\Delta }^{9})^{\frac{1}{2}}}e^{%
\frac{M_{\pi }^{2}b_{e}^{2}}{2}}\right] \frac{1}{3}\frac{e^{-M_{\pi }R}}{R}%
\,,
\end{equation}%
where the factor $160/9\sqrt{2}$ corresponds to $\left\langle N\Delta
\left\vert (\vec{\sigma}_{3}\cdot \vec{\sigma}_{6})(\vec{\tau}_{3}\cdot \vec{%
\tau}_{6})\right\vert NN\right\rangle $ for total spin and isospin equal to
1, multiplied by 9 the number of equivalent diagrams, and $b_{\mu },b_{+}$
and $b_{e}$ are defined as 
\begin{eqnarray}
\frac{1}{b_{\mu }^{2}} &=&\frac{1}{2b^{2}}+\frac{1}{2b_{\Delta }^{2}}  \notag
\\
\frac{1}{b_{\pm }^{2}} &=&\pm \frac{1}{b^{2}}+\frac{1}{b_{\mu }^{2}} \\
\frac{1}{b_{e}^{2}} &=&\frac{1}{2b_{+}^{2}}-\frac{b_{+}^{2}}{2b_{-}^{4}}\,. 
\notag
\end{eqnarray}%
Note that when $b_{\Delta }=b$ we also have $b_{e}=b_{\mu }=\sqrt{2}b_{+}=b$
hence reproducing, except for the spin-isospin factor, the $NN$ result.

At the baryonic level we have the lagrangians $\mathcal{L}_{\pi NN}^{PV}$ as
given by Eq.~(\ref{pinPV}) and 
\begin{equation}
\mathcal{L}_{\pi N\Delta }^{PV}=-\frac{(f_{\pi N\Delta })_{Q^{2}=M_{\pi
}^{2}}}{M_{\pi }}\,\,\bar{N}\,\,\vec{T}\partial \vec{\pi}\,\,\Delta \,,
\label{pind}
\end{equation}%
with a form factor $G_{\pi N\Delta }(Q^{2})$ which we shall choose as,
\begin{equation}
G_{\pi N\Delta }(Q^{2})\equiv \left( \frac{\Lambda _{\pi N\Delta
}^{2}-M_{\pi }^{2}}{\Lambda _{\pi N\Delta }^{2}-Q^{2}}\right) ^{\frac{1}{2}%
}\,,
\end{equation}%
In Eq.~(\ref{pind}) $\Delta $ corresponds to a Rarita-Schwinger $3/2$ spinor
field and $\vec{T}$ is the isospin nucleon-delta transition operator. The
corresponding pion exchange static central potential behaves asymptotically
as 
\begin{equation}
(\mathcal{V}_{B})_{\mathrm{OPE}}^{NN\rightarrow N\Delta }(R\rightarrow
\infty )=\frac{(f_{\pi NN})_{Q^{2}=M_{\pi }^{2}}(f_{\pi N\Delta
})_{Q^{2}=M_{\pi }^{2}}}{4\pi }\left[ \left\langle (\vec{\sigma}_{N}\cdot 
\vec{S})(\vec{\tau}_{N}\cdot \vec{T})\right\rangle \right] \frac{1}{3}\,%
\frac{e^{-M_{\pi }R}}{R}
\end{equation}%
where $\vec{S}$ is the spin nucleon-delta transition operator and $%
\left\langle (\vec{\sigma}_{N}\cdot \vec{S})(\vec{\tau}_{N}\cdot \vec{T}%
)\right\rangle =8/3$ for total spin and isospin equal to 1. Thus we
identify, 
\begin{equation}
\frac{(f_{\pi NN})_{Q^{2}=M_{\pi }^{2}}(f_{\pi N\Delta })_{Q^{2}=M_{\pi
}^{2}}}{4\pi }=\frac{f_{\pi qq}^{2}}{4\pi }\left[ \frac{20}{3\sqrt{2}}\,\,2%
\sqrt{2}\,\,\frac{b_{\mu }^{6}b_{+}^{3}b_{e}^{3}}{(b^{15}b_{\Delta }^{9})^{%
\frac{1}{2}}}\,\,e^{\frac{M_{\pi }^{2}b_{e}^{2}}{2}}\right] \,.
\end{equation}%
By using Eq.~(\ref{eqt}) we get, 
\begin{equation}
\frac{(f_{\pi N\Delta })_{Q^{2}=M_{\pi }^{2}}}{(f_{\pi NN})_{Q^{2}=M_{\pi
}^{2}}}=\left[ \frac{6\sqrt{2}}{5}\,\,2\sqrt{2}\,\,\frac{b_{\mu
}^{6}b_{+}^{3}b_{e}^{3}}{(b^{15}b_{\Delta }^{9})^{\frac{1}{2}}}\,\,e^{\frac{%
M_{\pi }^{2}(b_{e}^{2}-b^{2})}{2}}\right]
\end{equation}%
so that for $b_{\Delta }=b$ one obtains the usual spin-isospin relation ($%
f_{\pi N\Delta })_{Q^{2}=M_{\pi }^{2}}/(f_{\pi NN})_{Q^{2}=M_{\pi }^{2}}=6%
\sqrt{2}/5$.

For $b_{\Delta }$ in the interval $\left[ b,1.2b\right] $ we predict $%
(f_{\pi N\Delta }^{2})_{Q^{2}=M_{\pi }^{2}}/4\pi =\left[ 0.23,0.20\right] $
to be compared to $(f_{\pi N\Delta }^{2})_{Q^{2}=M_{\pi }^{2}}/4\pi \simeq
0.37,$ estimated from the $\Delta $ decay to $N\pi .$ This discrepancy seems
to confirm our initial expectations. If instead $f_{\pi qq}$ we had written
an effective ($f_{\pi qq})_{N\Delta }$ as a manner to take into account the $%
\Delta -N$ mass difference effect then the needed value to reproduce the
experimental number would have been $(f_{\pi qq})_{N\Delta }/f_{\pi qq}=%
\left[ 1.28,1.37\right] $ for $b_{\Delta }\in \left[ b,1.2b\right] $, i.e. $%
(f_{\pi qq})_{N\Delta }$ should be a $30\%$ bigger than $f_{\pi qq}$.

For nucleon resonances in general and in particular for $N^{\ast }(1440)$ we
expect the calculation of the coupling constants to be much more uncertain.
As a matter of fact the importance of relativistic corrections in the
description of the structure and decay of $N^{\ast }(1440)$ in terms of
quarks have been emphasized for a long time in the literature (see for
instance~\cite{Can96}). Furthermore the nature of the $N^{\ast }(1440)$ may
involve more than a simple $3q$ structure and the coupling of $q\overline{q}$
pairs to the meson structure can be relevant. Therefore the calculation of $%
\pi NN^{\ast }(1440)$ and $\sigma NN^{\ast }(1440)$ coupling constants
carried out in a preceding paper within the same framework~\cite{Bru02}
should be considered too simplistic. A less approximative calculation for
the pion case (involving also $\pi N\Delta $, $\pi N\Delta (1600)$ and $\pi
NN(1535))$ has been carried out in reference~\cite{Bru04} with Poincar\'{e}
covariant constituent quark models with instant, point and front forms of
relativistic kinematics; from the persistent deviation from data of the
calculated results the authors suggest the presence of sizable $qqqq%
\overline{q}$ components in the baryon wave functions.

\section{Light and strange quark sector: SU(3)$\mathbf{\otimes}$ SU(3).}

\label{secIII} The generalization of the chiral lagrangian to SU(3)$\otimes $%
SU(3) is straightforward. It is expressed as 
\begin{equation}
\tilde{\mathcal{L}}_{ch}=-\tilde{g}_{ch}\,\,\bar{q}%
\,\left( \sum_{a=0}^{8}\sigma _{a}\lambda _{a}+i\sum_{a=0}^{8}\gamma _{5}\pi
_{a}\lambda _{a}\right) \,q\,,  \label{eqg}
\end{equation}%
where $q$ has components $u,d$ and $s$, $\sigma _{0}$ and $\pi _{0}$ stand
for the scalar and pseudoscalar meson singlets whereas $\sigma _{i}$ and $%
\pi _{i}$ $(i=1...8)$ are the scalar and pseudoscalar meson octets.

In order to derive a potential involving the exchange of $\sigma ,$ the
SU(2) singlet, we shall assume the ideal mixing 
\begin{eqnarray}
\sigma _{0} &=&\sqrt{2/3}\,\sigma +\sqrt{1/3}\,(s\overline{s})  \notag \\
\sigma _{8} &=&\sqrt{1/3}\,\sigma -\sqrt{2/3}\,(s\overline{s})\,.
\end{eqnarray}%
When substituting these expressions in Eq.~(\ref{eqg}) the piece containing
the $\sigma $ and the $\vec{\pi}$ read, 
\begin{equation}
\tilde{\mathcal{L}}_{ch(\pi ,\sigma )}=-\tilde{g}_{ch%
}\,\,\bar{q}\,\left[ \sigma \left( \sqrt{2/3}\,\lambda _{0}+\sqrt{1/3}%
\,\lambda _{8}\right) +i\gamma _{5}\vec{\tau}\vec{\pi}\right] \,q
\label{pisigq}
\end{equation}%
where $\lambda _{0}\equiv \sqrt{2/3}\,\openone_{\mathrm{SU(3)}}$ and $%
\lambda _{8}\equiv \sqrt{3}\,Y$, being $Y$ the hypercharge. It is then clear
that for $q=u,d$ ($Yu=1/3=Yd)$ one formally recovers the SU(2)$\otimes $%
SU(2) lagrangian: $\openone_{\mathrm{SU(2)}}\equiv 2/3\,\openone_{\mathrm{%
SU(3)}}+Y$.

\noindent
Again we take into account SCSB through a vertex form factor $\tilde{F}%
(Q^{2})$ 
\begin{equation}
\tilde{F}(Q^{2})=\left( \frac{\tilde{\Lambda}^{2}}{\tilde{\Lambda}^{2}-Q^{2}}%
\right) ^{1/2}\,
\end{equation}%
Note also that the SCSB relation, $M_{\sigma }^{2}-M_{\pi }^{2}=4M_{q}^{2}$,
is preserved since it is derived for a non-strange $\sigma $ \cite{Sca82}.

\subsection{Parameters}

One should realize that the values of the couplings $\tilde{g}_{ch%
}\,\tilde{F}(Q^{2})$ (or equivalently the on-shell couplings $\tilde{g}%
_{\pi qq}$, $\tilde{g}_{\sigma qq}$ and the cutoff parameter $\tilde{\Lambda}
$) and the size parameter $\tilde{b}$ in SU(3)$\otimes $SU(3) need not be
the same as in SU(2)$\otimes $SU(2). This is easily understandable by
thinking for instance of the extra contribution to the $NN$ interaction
coming from $\eta $, $\eta ^{\prime }$ and $a_{0}$ in SU(3)$\otimes $SU(3).
This contribution is taken into account in an effective manner in SU(2)$%
\otimes $SU(2), where no $\eta $, $\eta ^{\prime }$, and $a_{0}$ are
present, through the fitted values of $g_{\pi qq}$, $g_{\sigma qq}$, $%
\Lambda $ and $b$. Fortunately we can correlate the variations of $b$ and $%
\Lambda $ (or $\tilde{b}$ and $\tilde{\Lambda})$ through the relation Eq.~(%
\ref{eqk}), 
\begin{equation}
\frac{(g_{\sigma NN}^{2})_{Q^{2}=M_{\sigma }^{2}}}{(g_{\pi
NN}^{2})_{Q^{2}=M_{\pi }^{2}}}=\frac{9}{25}\,\,\frac{\Lambda ^{2}-M_{\pi
}^{2}}{\Lambda ^{2}-M_{\sigma }^{2}}\,\,e^{\frac{(M_{\sigma }^{2}-M_{\pi
}^{2})b^{2}}{2}}=\frac{9}{25}\,\,\frac{\tilde{\Lambda}^{2}-M_{\pi }^{2}}{%
\tilde{\Lambda}^{2}-M_{\sigma }^{2}}\,\,e^{\frac{(M_{\sigma }^{2}-M_{\pi
}^{2})\tilde{b}^{2}}{2}}\,.
\end{equation}%
Then from the selected experimental value $(g_{\pi NN}^{2})_{Q^{2}=M_{\pi
}^{2}}/4\pi \simeq 14.6$ and from our prediction $(g_{\sigma
NN}^{2})_{Q^{2}=M_{\sigma }^{2}}/4\pi =68.2$ we get, for a typical value of $%
\tilde{\Lambda}\simeq 5.2$ fm$^{-1}$ ($\simeq $ 1.0 GeV) (note that it has
to be higher than any mass of the scalar or pseudoscalar meson octets) an
harmonic oscillator parameter $\tilde{b}\simeq 0.6$ fm. Then from equivalent
relations to Eqs.~(\ref{pinpiq}) and (\ref{gsigq}) we get 
\begin{equation}
\frac{\tilde{g}_{\pi qq}^{2}}{4\pi }=0.54
\end{equation}%
\begin{equation}
\frac{\tilde{g}_{\sigma qq}^{2}}{4\pi }=0.93\,,
\end{equation}%
and 
\begin{equation}
\tilde{g}_{ch}=g_{ch}\,.
\end{equation}%
For the sake of completeness we give the $B_{8}$ spatial wave function in
the SU(3)$_{\mathrm{flavor}}$ limit, 
\begin{equation}
(\Phi _{B_{8}})_{\mathrm{spatial}}(1,2,3;+\vec{R}/2)=\prod_{i=1}^{3}\left( 
\frac{1}{\pi \tilde{b}^{2}}\right) ^{\frac{3}{4}}\exp \left[ -\left( \vec{r}%
_{i}-\vec{R}/2\right) ^{2}/2\tilde{b}^{2}\right] \,.
\end{equation}

\subsection{$\mathbf{\protect\sigma B_{8}B_{8}}$}

\label{secIIIA} According to our preceding discussions the $\sigma qq$
lagrangian will be written as 
\begin{equation}
\tilde{\mathcal{L}}_{\sigma qq}=-\tilde{g}_{\sigma qq}\,\,\bar{q}\,%
\openone_{\mathrm{SU(2)}}\,\sigma \,\,q\, ,
\end{equation}%
with a vertex form factor 
\begin{equation}
\tilde{F}_{\sigma qq}(Q^{2})\equiv \left( \frac{\tilde{\Lambda}%
^{2}-M_{\sigma }^{2}}{\tilde{\Lambda}^{2}-Q^{2}}\right) ^{1/2} \, .
\end{equation}%
\begin{figure}[th]
\vspace*{-2cm}
\epsfig{file=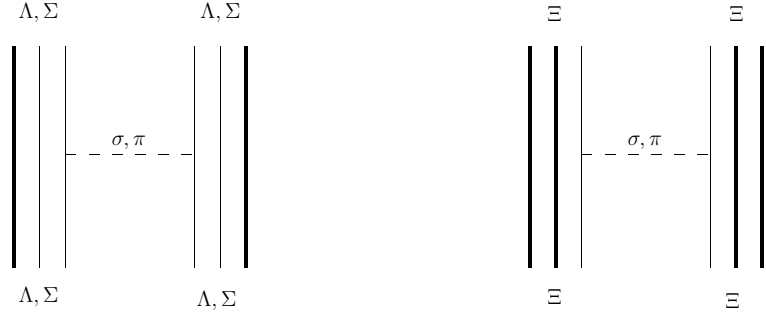,width=12cm}
\vspace*{-9cm}
\caption{Asymptotic $\Lambda \Lambda $, $\Sigma \Sigma $, and $\Xi \Xi $ OPE
and OSE interactions at quark level. Thin lines stand for light ($u,d$)
quarks and thick lines for strange ($s$) quarks.}
\label{f4}
\end{figure}
From this lagrangian it is clear that the only difference when calculating
the asymptotic potential at the quark level for the several $B_{8}B_{8}$'s
has to do with the number of the pairs of light quarks ($u,d)$ in them
allowing for the exchange of the $\sigma $, i.e., with the number of
equivalent diagrams, Figs.~\ref{f1} and \ref{f4}. This number is 9 for $NN,$
4 for $\Lambda \Lambda $ and $\Sigma \Sigma $, and 1 for $\Xi \Xi $. Thus 
\begin{eqnarray}
(\tilde{V}_{q})_{\mathrm{OSE}}^{\Lambda \Lambda \rightarrow \Lambda
\Lambda }(R\rightarrow \infty ) &=&(\tilde{V}_{q})_{\mathrm{OSE}%
}^{\Sigma \Sigma \rightarrow \Sigma \Sigma }(R\rightarrow \infty )=\frac{4}{9%
}(\tilde{V}_{q})_{\mathrm{OSE}}^{NN\rightarrow NN}(R\rightarrow \infty )
\notag \\
(\tilde{V}_{q})_{\mathrm{OSE}}^{\Xi \Xi \rightarrow \Xi \Xi
}(R\rightarrow \infty ) &=&\frac{1}{9}(\tilde{V}_{q})_{\mathrm{OSE}%
}^{NN\rightarrow NN}(R\rightarrow \infty )\,.
\end{eqnarray}%
On the other hand at the baryonic level we shall write the lagrangian as 
\begin{equation}
\mathcal{L}_{\sigma B_{8}B_{8}}\equiv -(g_{\sigma
B_{8}B_{8}})_{Q^{2}=M_{\sigma }^{2}}\,\,\bar{B_{8}}\,\sigma \,\,B_{8}\,,
\label{eqna2}
\end{equation}%
with a vertex form factor 
\begin{equation}
G_{\sigma B_{8}B_{8}}(Q^{2})=\left( \frac{\Lambda _{\sigma
B_{8}B_{8}}^{2}-M_{\sigma }^{2}}{{\Lambda _{\sigma B_{8}B_{8}}^{2}-Q^{2}}}%
\right) ^{1/2}\,.
\end{equation}%
so that $G_{\sigma B_{8}B_{8}}(Q^{2}=M_{\sigma }^{2})=1$ and where the $%
(g_{\sigma B_{8}B_{8}})_{Q^{2}=M_{\sigma }^{2}}\,$ are expressed in
conventional notation 
\begin{eqnarray}
(g_{\sigma NN})_{Q^{2}=M_{\sigma }^{2}}\, &\equiv &\sqrt{2/3}%
\,g_{s,1}+1/3\,g_{s,+}\,(4\alpha _{s}-1)  \notag \\
(g_{\sigma \Lambda \Lambda })_{Q^{2}=M_{\sigma }^{2}}\, &\equiv &\sqrt{2/3}%
\,g_{s,1}-2/3\,g_{s,+}\,(1-\alpha _{s})  \notag \\
(g_{\sigma \Sigma \Sigma })_{Q^{2}=M_{\sigma }^{2}}\, &\equiv &\sqrt{2/3}%
g_{s,1}+2/3\,g_{s,+}\,(1-\alpha _{s})  \label{alfas} \\
(g_{\sigma \Xi \Xi })_{Q^{2}=M_{\sigma }^{2}}\, &\equiv &\sqrt{2/3}%
\,g_{s,1}-1/3\,g_{s,+}\,(1+2\alpha _{s})\,,  \notag
\end{eqnarray}%
being $g_{s,1}\equiv (g_{\sigma _{0}B_{8}B_{8}})_{Q^{2}=M_{\sigma }^{2}}$
the scalar SU(3) singlet coupling constant, $g_{s,+}\equiv g_{s,D}+g_{s,F}$
the sum of the $D$ (symmetric) and the $F$ (antisymmetric) scalar coupling
constants in SU(3) and $\alpha _{s}\equiv g_{s,F}/g_{s,+}$ the $F/(F+D)$
ratio of the scalar octet.

From this baryonic lagrangian the following relations between the asymptotic
potentials come out 
\begin{eqnarray}
(\mathcal{V}_{B})_{\mathrm{OSE}}^{\Lambda \Lambda \rightarrow \Lambda
\Lambda }\,(R\rightarrow \infty ) &=&\frac{(g_{\sigma \Lambda \Lambda
}^{2})_{Q^{2}=M_{\sigma }^{2}}}{(g_{\sigma NN}^{2})_{Q^{2}=M_{\sigma }^{2}}}(%
\mathcal{V}_{B})_{\mathrm{OSE}}^{NN\rightarrow NN}\,(R\rightarrow \infty ) 
\notag \\
(\mathcal{V}_{B})_{\mathrm{OSE}}^{\Sigma \Sigma \rightarrow \Sigma \Sigma
}\,(R\rightarrow \infty ) &=&\frac{(g_{\sigma \Sigma \Sigma
}^{2})_{Q^{2}=M_{\sigma }^{2}}}{(g_{\sigma NN}^{2})_{Q^{2}=M_{\sigma }^{2}}}(%
\mathcal{V}_{B})_{\mathrm{OSE}}^{NN\rightarrow NN}\,(R\rightarrow \infty ) \\
(\mathcal{V}_{B})_{\mathrm{OSE}}^{\Xi \Xi \rightarrow \Xi \Xi
}\,(R\rightarrow \infty ) &=&\frac{(g_{\sigma \Xi \Xi
}^{2})_{Q^{2}=M_{\sigma }^{2}}}{(g_{\sigma NN}^{2})_{Q^{2}=M_{\sigma }^{2}}}(%
\mathcal{V}_{B})_{\mathrm{OSE}}^{NN\rightarrow NN}\,(R\rightarrow \infty )\,.
\notag  \label{eqnb}
\end{eqnarray}%
From the comparison of the asymptotic potentials at the quark and baryon
levels we immediately get relations between the coupling constants%
\begin{eqnarray}
\frac{(g_{\sigma \Lambda \Lambda }^{2})_{Q^{2}=M_{\sigma }^{2}}}{(g_{\sigma
NN}^{2})_{Q^{2}=M_{\sigma }^{2}}} &=&\frac{(g_{\sigma \Sigma \Sigma
}^{2})_{Q^{2}=M_{\sigma }^{2}}}{(g_{\sigma NN}^{2})_{Q^{2}=M_{\sigma }^{2}}}=%
\frac{4}{9}  \notag \\
\frac{(g_{\sigma \Xi \Xi }^{2})_{Q^{2}=M_{\sigma }^{2}}}{(g_{\sigma
NN}^{2})_{Q^{2}=M_{\sigma }^{2}}} &=&\frac{1}{9}\,.  \label{eqr}
\end{eqnarray}%
We should emphasize that these ratios are preserved in the limit $M_{\sigma
}\rightarrow 0$, i.e., 
\begin{eqnarray}
\frac{(g_{\sigma \Lambda \Lambda }^{2})_{Q^{2}=M_{\sigma }^{2}\rightarrow 0}%
}{(g_{\sigma NN}^{2})_{Q^{2}=M_{\sigma }^{2}\rightarrow 0}} &=&\frac{%
(g_{\sigma \Sigma \Sigma }^{2})_{Q^{2}=M_{\sigma }^{2}\rightarrow 0}}{%
(g_{\sigma NN}^{2})_{Q^{2}=M_{\sigma }^{2}\rightarrow 0}}=\frac{4}{9}  \notag
\\
\frac{(g_{\sigma \Xi \Xi }^{2})_{Q^{2}=M_{\sigma }^{2}\rightarrow 0}}{%
(g_{\sigma NN}^{2})_{Q^{2}=M_{\sigma }^{2}\rightarrow 0}} &=&\frac{1}{9}\,.
\label{eqr0}
\end{eqnarray}%
It is interesting to compare these $M_{\sigma }\rightarrow 0$ ratios with
the average QCDSR predictions at $Q^{2}=0$ in the SU(3) limit. From Ref.~%
\cite{Oka06} we take 
\begin{eqnarray}
\frac{(g_{\sigma \Lambda \Lambda }^{2})_{Q^{2}=0}}{(g_{\sigma
NN}^{2})_{Q^{2}=0}} &=&(0.43)^{2}=0.19  \notag \\
\frac{(g_{\sigma \Sigma \Sigma }^{2})_{Q^{2}=0}}{(g_{\sigma
NN}^{2})_{Q^{2}=0}} &=&(0.91)^{2}=0.83  \notag \\
\frac{(g_{\sigma \Xi \Xi }^{2})_{Q^{2}=0}}{(g_{\sigma NN}^{2})_{Q^{2}=0}}
&=&(0.08)^{2}=0.006.  \label{ratioka}
\end{eqnarray}%
As can be checked our ratios differ by a factor $2$ ($1/2)$ for $\Lambda -N$
and $\Sigma -N$ ($\Xi -N)$ from the QCDSR ones. We may interpret this again
as a reflection of the non-Goldstone character of the $\sigma $ meson.

\subsubsection{Coupling constants and form factors values}

From Eq.~(\ref{eqr}) and from the calculated value $(g_{\sigma
NN}^{2})_{Q^{2}=M_{\sigma }^{2}}/4\pi =68.2$ we predict the coupling
constants 
\begin{eqnarray}
\frac{(g_{\sigma \Lambda \Lambda }^{2})_{Q^{2}=M_{\sigma }^{2}}}{4\pi } &=&%
\frac{(g_{\sigma \Sigma \Sigma }^{2})_{Q^{2}=M_{\sigma }^{2}}}{4\pi }=30.3 
\notag \\
\frac{(g_{\sigma \Xi \Xi }^{2})_{Q^{2}=M_{\sigma }^{2}}}{4\pi } &=&7.6\,.
\end{eqnarray}%
Concerning the $F/(F+D)$ ratio we have from Eq.~(\ref{eqr}) $(g_{\sigma
\Lambda \Lambda })_{Q^{2}=M_{\sigma }^{2}}=(g_{\sigma \Sigma \Sigma
})_{Q^{2}=M_{\sigma }^{2}}$ what implies from Eq.~(\ref{alfas}) 
\begin{equation}
\alpha _{s}=1\,,  \label{alfaval}
\end{equation}%
or 
\begin{equation}
g_{S,D}=0.  \label{gs0}
\end{equation}%
Regarding $g_{s,1}$ and $g_{S,+}$ we can use the fact that from the ideal
mixing we have assumed $g_{(s\overline{s})NN}=-1/\sqrt{3}\,g_{s,1}+\sqrt{2}%
\,g_{s,F}-\sqrt{2}/3\,g_{s,D}=0$. Then from Eq.~(\ref{gs0}) we immediately
obtain $g_{s,+}=1/\sqrt{6}\,g_{s,1}.$ If we substitute this relation and
Eq.~(\ref{alfaval}) in the first expression of Eq.~(\ref{alfas}) we get $%
(g_{\sigma NN})_{Q^{2}=M_{\sigma }^{2}}=3g_{s,+}$ from where 
\begin{eqnarray}
\frac{g_{s,+}^{2}}{4\pi } &=&7.6  \label{gsmas} \\
\frac{g_{s,1}^{2}}{4\pi } &=&45.5\,,
\end{eqnarray}%
satisfying 
\begin{equation}
\dfrac{g_{s,1}^{2}}{g_{s,+}^{2}}=6\,.
\end{equation}%
With respect to the cutoff parameters $\Lambda _{\sigma B_{8}B_{8}}$ we can
tentatively use our on-shell couplings ratios, Eq.~(\ref{eqr}), altogether
with the QCDSR ones at $Q^{2}=0$ detailed above, Eq.~(\ref{ratioka}), to
establish a range of variation for them. Explicitly we can write%
\begin{equation}
\left( \frac{(g_{\sigma B_{8}B_{8}}^{2})_{Q^{2}=0}}{(g_{\sigma
NN}^{2})_{Q^{2}=0}}\right) =\left( \frac{(g_{\sigma
B_{8}B_{8}}^{2})_{Q^{2}=M_{\sigma }^{2}}}{(g_{\sigma
NN}^{2})_{Q^{2}=M_{\sigma }^{2}}}\right) \frac{\left( \frac{\Lambda _{\sigma
B_{8}B_{8}}^{2}-M_{\sigma }^{2}}{{\Lambda _{\sigma B_{8}B_{8}}^{2}}}\right) 
}{\left( \frac{\Lambda _{\sigma NN}^{2}-M_{\sigma }^{2}}{\Lambda _{\sigma
NN}^{2}}\right) }\,,
\end{equation}%
or equivalently 
\begin{equation}
\Lambda _{\sigma B_{8}B_{8}}=\Lambda _{\sigma NN}\left( \frac{M_{\sigma }^{2}%
}{\Lambda _{\sigma NN}^{2}(1-x_{B_{8}})+x_{B_{8}}M_{\sigma }^{2}}\right) ^{%
\frac{1}{2}}\,,
\end{equation}%
where 
\begin{equation}
x_{B_{8}}\equiv \frac{\left( \frac{(g_{\sigma B_{8}B_{8}}^{2})_{Q^{2}=0}}{%
(g_{\sigma NN}^{2})_{Q^{2}=0}}\right) }{\left( \frac{(g_{\sigma
B_{8}B_{8}}^{2})_{Q^{2}=M_{\sigma }^{2}}}{(g_{\sigma
NN}^{2})_{Q^{2}=M_{\sigma }^{2}}}\right) }\,.
\end{equation}%
By using the average value $\Lambda _{\sigma NN}=3.97$ fm$^{-1}$, Eq.~(\ref%
{cutoff}), we get 
\begin{eqnarray}
\Lambda _{\sigma NN} &=&3.97\,\mathrm{{fm}^{-1}}  \notag \\
\Lambda _{\sigma \Lambda \Lambda } &=&4.48\,\mathrm{{fm}^{-1}}  \notag \\
\Lambda _{\sigma \Sigma \Sigma } &=&4.75\,\mathrm{{fm}^{-1}} \\
\Lambda _{\sigma \Xi \Xi } &=&3.45\,\mathrm{{fm}^{-1}\,}\,.  \notag
\end{eqnarray}%
Thus we can use $\Lambda _{\sigma B_{8}B_{8}}\simeq 4.0\,\mathrm{{fm}^{-1}}$
as an average value for the whole baryon octet.

\subsection{$a_0\mathbf{B_{8}B_{8}}$}

\label{secIIIB} The results obtained for $\sigma $ can be extrapolated to
the $a_{0}$ meson in a straightforward way. Let us recall that in the
additive constituent quark pattern one has degenerate masses $%
M_{a_{0}}=M_{\sigma }$ (see Ref.~\cite{Sca82} for an explanation of the
non--degeneracy). Then by writing the $a_{0}$ on--shell couplings in SU(3)
language 
\begin{eqnarray}
(g_{a_{0}NN})_{Q^{2}=M_{a_{0}}^{2}} &=&g_{s,+}  \notag \\
(g_{a_{0}\Sigma \Sigma })_{Q^{2}=M_{a_{0}}^{2}} &=&2g_{s,+}\alpha _{s} \\
(g_{a_{0}\Xi \Xi })_{Q^{2}=M_{a_{0}}^{2}} &=&g_{s,+}(2\alpha _{s}-1)\,, 
\notag
\end{eqnarray}%
and imposing the degeneracy we predict from our value for $g_{s,+}$ (Eq.~(%
\ref{gsmas})) 
\begin{eqnarray}
\dfrac{(g_{a_{0}NN}^{2})_{Q^{2}=M_{a_{0}}^{2}}}{4\pi } &=&\dfrac{%
(g_{a_{0}\Xi \Xi }^{2})_{Q^{2}=M_{a_{0}}^{2}}}{4\pi }=7.6  \notag \\
\dfrac{(g_{a_{0}\Sigma \Sigma }^{2})_{Q^{2}=M_{a_{0}}^{2}}}{4\pi } &=&30.3\,.
\end{eqnarray}

\subsection{$\mathbf{\protect\pi B_{8}B_{8}}$}

\label{secIIIC} In order to avoid mass factors we shall use the pseudovector
coupling for the pion, i.e., the lagrangian 
\begin{equation}
\tilde{\mathcal{L}}_{\pi qq}^{PV}=-\frac{\tilde{f}_{\pi qq}}{M_{\pi }%
}\bar{q}\,\,i\gamma _{5}\gamma _{\mu }\vec{\tau}\partial ^{\mu }\vec{\pi}%
\,\,q
\end{equation}%
with the vertex form factor 
\begin{equation}
\tilde{F}_{\pi qq}(Q^{2})\equiv \left( \frac{\tilde{\Lambda}^{2}-M_{\pi }^{2}%
}{\tilde{\Lambda}^{2}-Q^{2}}\right) ^{1/2}
\end{equation}%
to derive the asymptotic pion exchange central potentials. To perform the
calculation we choose a total (spin, isospin) in each case. By considering
the $^{1}S_{0}$ partial wave for instance we take $(0,1)$ for $NN$ and $\Xi
\Xi $ and $(0,0)$ for $\Sigma \Sigma $ interactions. Then we have 
\begin{eqnarray}
(V_{q})_{\mathrm{OPE}}^{NN\rightarrow NN}\,(R\rightarrow \infty ) &=&\frac{%
\tilde{f}_{\pi qq}^{2}}{4\pi }\left[ -\frac{75}{9}\,e^{\frac{M_{\pi }^{2}%
\tilde{b}^{2}}{2}}\right] \frac{1}{3}\,\frac{e^{-M_{\pi }R}}{R}  \notag \\
(V_{q})_{\mathrm{OPE}}^{\Sigma \Sigma \rightarrow \Sigma \Sigma
}\,(R\rightarrow \infty ) &=&\frac{\tilde{f}_{\pi qq}^{2}}{4\pi }\left[ 
\frac{32}{3}\,e^{\frac{M_{\pi }^{2}\tilde{b}^{2}}{2}}\right] \frac{1}{3}\,%
\frac{e^{-M_{\pi }R}}{R} \\
(V_{q})_{\mathrm{OPE}}^{\Xi \Xi \rightarrow \Xi \Xi }\,(R\rightarrow \infty
) &=&\frac{\tilde{f}_{\pi qq}^{2}}{4\pi }\left[ -\frac{3}{9}\,e^{\frac{%
M_{\pi }^{2}\tilde{b}^{2}}{2}}\right] \frac{1}{3}\,\frac{e^{-M_{\pi }R}}{R}%
\,.  \notag
\end{eqnarray}%
On the other hand at the baryonic level the lagrangians can be expressed as 
\begin{equation}
\mathcal{L}_{\pi NN}^{PV}=-\frac{(f_{\pi NN})_{Q^{2}=M_{\pi }^{2}}}{M_{\pi }}%
\,\bar{N}\,i\gamma _{5}\gamma _{\mu }\,\vec{\tau}\,\partial ^{\mu }\vec{\pi}%
\,\,N
\end{equation}%
\begin{equation}
\mathcal{L}_{\pi \Sigma \Sigma}^{PV}=-\frac{(f_{\pi \Sigma \Sigma })_{Q^{2}=M_{\pi
}^{2}}}{M_{\pi }}(\vec{\bar{\Sigma}}\times \vec{\Sigma}\,)\gamma
_{5}\,\gamma _{\mu }\,\partial ^{\mu }\vec{\pi}
\end{equation}%
\begin{equation}
\mathcal{L}_{\pi \Xi \Xi}^{PV}=-\frac{(f_{\pi \Xi \Xi })_{Q^{2}=M_{\pi }^{2}}}{%
M_{\pi }}\,\bar{\Xi}\,i\gamma _{5}\,\gamma _{\mu }\,\vec{\tau}\,\partial
^{\mu }\,\vec{\pi}\,\Xi \, ,
\end{equation}%
with form factors 
\begin{eqnarray}
G_{\pi NN}(Q^{2}) &=&\left( \frac{\Lambda _{\pi NN}^{2}-M_{\pi }^{2}}{%
\Lambda _{\pi NN}^{2}-Q^{2}}\right) ^{\frac{1}{2}}  \notag \\
G_{\pi \Sigma \Sigma }(Q^{2}) &=&\left( \frac{\Lambda _{\pi \Sigma \Sigma
}^{2}-M_{\pi }^{2}}{\Lambda _{\pi \Sigma \Sigma }^{2}-Q^{2}}\right) ^{\frac{1%
}{2}}  \label{alfap} \\
G_{\pi \Xi \Xi }(Q^{2}) &=&\left( \frac{\Lambda _{\pi \Xi \Xi }^{2}-M_{\pi
}^{2}}{\Lambda _{\pi \Xi \Xi }^{2}-Q^{2}}\right) ^{\frac{1}{2}}  \, ,
\end{eqnarray}%
and where in conventional SU(3) notation 
\begin{eqnarray*}
(f_{\pi NN})_{Q^{2}=M_{\pi }^{2}} &=&f_{p,+} \\
(f_{\pi \Sigma \Sigma })_{Q^{2}=M_{\pi }^{2}} &=&2f_{p,+}\, \alpha _{p} \\
(f_{\pi \Xi \Xi })_{Q^{2}=M_{\pi }^{2}} &=&-f_{p,+}(1-2\alpha _{p})\,,
\end{eqnarray*}%
having introduced $f_{p,+}\equiv (f_{p,D}+f_{p,F})$ as the sum of the
symmetric and antisymmetric pseudoscalar octet coupling constants, and $%
\alpha _{p}\equiv f_{P,F}/(f_{P,D}+f_{P,F})$ as the $F/(F+D)$ ratio of the
pseudoscalar octet$.$

By using the same total (spin, isospin) channels as above the corresponding
asymptotic pion exchange central potentials are 
\begin{eqnarray}
(\mathcal{V}_{B})_{\mathrm{OPE}}^{NN\rightarrow NN}\,(R\rightarrow \infty )
&=&\frac{(f_{\pi NN}^{2})_{Q^{2}=M_{\pi }^{2}}}{4\pi }\left[ -3\right] \frac{%
1}{3}\frac{e^{-M_{\pi }R}}{R}  \notag \\
(\mathcal{V}_{B})_{\mathrm{OPE}}^{\Sigma \Sigma \rightarrow \Sigma \Sigma
}\,(R\rightarrow \infty ) &=&\frac{(f_{\pi \Sigma \Sigma
}^{2})_{Q^{2}=M_{\pi }^{2}}}{4\pi }\left[ 6\right] \frac{1}{3}\frac{%
e^{-M_{\pi }R}}{R} \\
(\mathcal{V}_{B})_{\mathrm{OPE}}^{\Xi \Xi \rightarrow \Xi \Xi
}\,(R\rightarrow \infty ) &=&\frac{(f_{\pi \Xi \Xi }^{2})_{Q^{2}=M_{\pi
}^{2}}}{4\pi }\left[ -3\right] \frac{1}{3}\frac{e^{-M_{\pi }R}}{R}\,.  \notag
\end{eqnarray}%
Then from the comparison of the asymptotic potentials at the baryon level
and quark level the following relations come out 
\begin{eqnarray}
(f_{\pi NN}^{2})_{Q^{2}=M_{\pi }^{2}} &=&\tilde{f}_{\pi qq}^{2}\,\frac{25}{9}%
\,e^{\frac{M_{\pi }^{2}\tilde{b}^{2}}{2}}  \notag \\
(f_{\pi \Sigma \Sigma }^{2})_{Q^{2}=M_{\pi }^{2}} &=&\tilde{f}_{\pi qq}^{2}\,%
\frac{16}{9}\,e^{\frac{M_{\pi }^{2}\tilde{b}^{2}}{2}} \\
(f_{\pi \Xi \Xi }^{2})_{Q^{2}=M_{\pi }^{2}} &=&\tilde{f}_{\pi qq}^{2}\,\frac{%
1}{9}\,e^{\frac{M_{\pi }^{2}\tilde{b}^{2}}{2}}\,,  \notag
\end{eqnarray}%
and 
\begin{eqnarray}
\frac{(f_{\pi \Sigma \Sigma }^{2})_{Q^{2}=M_{\pi }^{2}}}{(f_{\pi
NN}^{2})_{Q^{2}=M_{\pi }^{2}}} &=&\frac{16}{25}  \notag \\
\frac{(f_{\pi \Xi \Xi }^{2})_{Q^{2}=M_{\pi }^{2}}}{(f_{\pi
NN}^{2})_{Q^{2}=M_{\pi }^{2}}} &=&\frac{1}{25}\,.  \label{su3}
\end{eqnarray}

\subsubsection{Coupling constants and form factors values}

From Eqs.~(\ref{su3}) and the standard value $(f_{\pi NN}^{2})_{Q^{2}=M_{\pi
}^{2}}/4\pi =0.079$\ we predict the numerical values 
\begin{eqnarray}
\frac{(f_{\pi \Sigma \Sigma }^{2})_{Q^{2}=M_{\pi }^{2}}}{4\pi } &=&0.051 
\notag \\
\frac{(f_{\pi \Xi \Xi }^{2})_{Q^{2}=M_{\pi }^{2}}}{4\pi } &=&0.0032\,,
\end{eqnarray}%
or 
\begin{eqnarray}
\frac{(g_{\pi \Sigma \Sigma }^{2})_{Q^{2}=M_{\pi }^{2}}}{4\pi } &=&9.4 
\notag \\
\frac{(g_{\pi \Xi \Xi }^{2})_{Q^{2}=M_{\pi }^{2}}}{4\pi } &=&0.6\,,
\end{eqnarray}%
and 
\begin{equation}
\alpha _{p}=0.4\,.
\end{equation}%
This compares quite well with a derived value of $\alpha _{p}\simeq 0.365\pm
0.007$ from the $F/D$ ratio extracted from semileptonic decays of baryons~%
\cite{Clo93}.

With respect to the couplings at $Q^{2}=0$ we can rely on the quite
approximate Goldstone boson character of the pion and assume they are given
by $(f_{\pi B_{8}B_{8}}^{2})_{Q^{2}=0}\simeq (f_{\pi
B_{8}B_{8}}^{2})_{Q^{2}=M_{\pi }^{2}\rightarrow 0}.$ As in this limit the
ratios between the couplings are the same than the on-shell ones obtained
above, Eq.~(\ref{su3}), we can use them altogether with $(f_{\pi
NN}^{2})_{Q^{2}=M_{\pi }^{2}\rightarrow 0}/4\pi =0.072$, Eq.~(\ref{gchiral}%
), to predict 
\begin{eqnarray}
\frac{(f_{\pi \Sigma \Sigma }^{2})_{Q^{2}=0}}{4\pi } &\simeq &0.046  \notag
\\
\frac{(f_{\pi \Xi \Xi }^{2})_{Q^{2}=0}}{4\pi } &\simeq &0.0029\,,
\end{eqnarray}%
or 
\begin{eqnarray}
\frac{(g_{\pi \Sigma \Sigma }^{2})_{Q^{2}=0}}{4\pi } &\simeq &8.5  \notag \\
\frac{(g_{\pi \Xi \Xi }^{2})_{Q^{2}=0}}{4\pi } &\simeq &0.5\, .
\end{eqnarray}
Moreover the preservation of the ratios implies the equality of the form
factors. From $(f_{\pi NN}^{2})_{Q^{2}=0}\simeq (f_{\pi
NN}^{2})_{Q^{2}=M_{\pi }^{2}\rightarrow 0}$ and $(f_{\pi
NN}^{2})_{Q^{2}=M_{\pi }^{2}}$ we deduce 
\begin{equation}
\Lambda _{\pi B_{8}B_{8}}\simeq 2.35 \,\, \mathrm{fm}^{-1}.
\end{equation}

\section{Summary}

\label{secIV} From a SU(2)$\otimes $SU(2) chiral quark lagrangian
incorporating spontaneous chiral symmetry breaking, and its generalization
to SU(3)$\otimes $SU(3), asymptotic meson exchange $B_{8}B_{8}\rightarrow
B_{8}B_{8}$ interaction potentials are derived in the Born-Oppenheimer
approximation. The comparison with the corresponding potentials from a SU(2)
or SU(3) invariant hadronic lagrangian allows for the expression of the $\pi
B_{8}B_{8}$ and $\sigma B_{8}B_{8}$ coupling constants in terms of the
elementary $\pi qq$ and $\sigma qq$ ones. By using the $\pi NN$ coupling
constant as an input the $\pi qq$ one gets fixed. From it the rest of
coupling constants ($\sigma qq,$ $\pi B_{8}B_{8}$ and $\sigma B_{8}B_{8})$
are predicted. Their $M_{m}\rightarrow 0$ limits are also of interest to be
compared with the values of the couplings at $Q^{2}=0$ provided by
phenomenological analyses or QCDSR. The similar value obtained for $\pi NN$
indicates the quite approximate Goldstone boson nature of the pion. On the
contrary $\sigma B_{8}B_{8}$ couplings are significantly different as might
be expected.

Further information about the couplings can be extracted under the
assumption that our on-shell model predictions and the $Q^{2}=0$ values from
external analyses can be managed jointly. Though this assumption is
debatable it allows to get some insight into the cutoff parameters $\Lambda
_{mB_{8}B_{8}}$ at the baryonic level.

\begin{table}[th]
\caption{Pion and sigma coupling constants to quarks in our
SU(2)$\otimes$SU(2) and SU(3)$\otimes$SU(3) models.}
\label{t2}
\begin{center}
\begin{tabular}{ccccc}
\hline\hline
$m$ & & SU(2)$\otimes$SU(2) & & SU(3)$\otimes$SU(3) \\ \hline
$(\pi,\sigma)$ & & $g^2_{ch}/4\pi=$0.535       & & $\tilde{g}^2_{ch}/4\pi=$0.535 \\
$\pi$          & & $g^2_{\pi qq}/4\pi=$0.55    & & $\tilde{g}^2_{\pi qq}/4\pi=$0.54 \\
$\sigma$       & & $g^2_{\sigma qq}/4\pi=$1.59 & & $\tilde{g}^2_{\sigma qq}/4\pi=$0.93 \\
\hline\hline
\end{tabular}%
\end{center}
\end{table}

We summarize in Table~\ref{t2} the pion and sigma coupling constants to
quarks in our models. In Table~\ref{t3} the values obtained for the coupling
constants and the form factors parameters at the baryon level are listed. 
\begin{table}[th]
\caption{Predicted pairs $\left( (g_{mB_{8}B_{8}}^{2}/4\protect\pi %
)_{Q^{2}=M_{m}^{2}}\text{ },\text{ }\Lambda _{mB_{8}B_{8}}(fm^{-1})\right) $
from the chiral quark potential model, for exact SU(3) symmetry ($M_{\Lambda
}=M_{\Sigma }=M_{\Xi }=M_{N}=939$ MeV). The superindex * indicates the $%
\protect\pi NN$ coupling constant value used as input.}
\label{t3}
\begin{center}
\begin{tabular}{rcccc}
\hline\hline
$m$ & $mNN$ & $m\Lambda \Lambda $ & $m\Sigma \Sigma $ & $m\Xi \Xi $ \\ \hline
$\pi $ & (14.6$^{\ast }$ $,$ 2.35) &  & (9.4 $,$ 2.35) & (0.6 $,$ 2.35) \\ 
$\sigma $ & (68.2 $,$ 3.97) & (30.3 $,$ 4.48) & (30.3 $,$ 4.75) & (7.6 $,$
3.45) \\ \hline\hline
\end{tabular}%
\end{center}
\end{table}
By making use of the $a_{0}-\sigma $ degeneracy in our quark model we have
also predicted $a_{0}B_{8}B_{8}$ on--shell couplings. Concerning other
diagonal $mBB$ couplings such as $f_{0}B_{8}B_{8}$, $\eta B_{8}B_{8}$ and $%
\eta ^{\prime }B_{8}B_{8}$ to which our formalism could be also applied the
situation gets complicated by the presence of the strange quark and/or
antiquark which may give rise to relevant SU(3) breaking effects out of the
scope of our symmetry treatment.

Let us finally add that in our model Goldberger-Treiman relations of the
form ($g_{A})_{\pi B_{8}B_{8}}/2f_{\pi }=(f_{\pi B_{8}B_{8}})_{Q^{2}=M_{\pi
}^{2}}/M_{\pi }$, where $g_{A}$ stands for the axial coupling constant and 
$f_{\pi }$ for the pion decay constant, can be immediately applied. In our
non-relativistic description $(g_{A})_{\pi NN}=5/3$, a value considerable 
larger than the experimental one $1.267.$ Consequently $f_{\pi }=116$ MeV, 
which is $20\%$ bigger than the experimental value of $93$ MeV. Regarding 
these discrepancies it has been shown~\cite{Bru04} that a relativistic 
treatment, beyond our present approach, could correct them to a good extent.

\acknowledgments This work has been partially funded by Ministerio de
Educaci\'on y Ciencia and EU FEDER under Contract No. FPA2007-65748,
by Junta de Castilla y Le\'{o}n under Contract No. SA016A17, and by the
Spanish Consolider-Ingenio 2010 Program CPAN (CSD2007-00042).

\end{document}